\documentstyle[aps]{revtex}

\begin{document}
\draft

\title{Influences of  broken time-reversal symmetry 
on the d.c. Josephson effects in $d$-wave superconductors }

\author{Yukio Tanaka }

\address{Graduate School of Science and Technology and
Department of Physics,  Niigata University, Ikarashi, Niigata 950-21,
Japan}

\author{Satoshi Kashiwaya$^{+}$ }

\address{Ginzton Laboratory, Stanford University,
Stanford, CA 94305-4085, USA}
\date{today}
\maketitle
\begin{abstract}
In order to examine the influences of the
spatial dependence of the pair potential,
the d.c. Josephson current in $d$-wave superconductors is calculated
using self-consistently determined pair potentials.
The results show that
the suppression of the $d$-wave pair potential near the insulator
does not have serious effect on the
properties of the Josephson current.
On the other hand, drastic changes are obtained due to the inducement of
a subdominant $s$-wave component, 
which spontaneously breaks time reversal symmetry.
Especially, a rapid enhancement of the Josephson current at low temperature
predicted in previous formulas is strongly suppressed.
\end{abstract}
\vspace{20pt}

\pacs{}
\widetext
Nowadays to clarify the properties of the 
tunneling junction in $d$-wave superconductors is an intriguing problem, 
since there are several evidences which support the 
$d$-wave symmetry of the pair potentials in
high-$T_{C}$ superconductors $^{1}$. 
The observation of the $\pi$-junction using Josephson junctions
serves as a very convincing evidence
for the $d$-wave symmetry of high-$T_{C}$ superconductors $^{2}$.
Moreover, recent theories have shown 
the existence of a novel interference effect in
the quasiparticle tunneling $^{3-5}$.
On the surfaces of $d$-wave superconductors,
zero energy states (ZES) is formed at the surfaces
as the results of sign change in the internal phases
of the pair potential
depending on the orientation of the surface.
The ZES is detectable by tunneling spectroscopy
as conductance peaks.
By extending this concept to the general anisotropic superconductors,
the phase-sensitivity of the tunneling spectroscopy has been predicted.
The formation of the ZES at the surfaces of high-$T_{c}$
superconductors
have been actually observed as zero energy peaks (ZEP)
in several tunneling spectroscopy experiments $^{5-8}$.
\par
As for the d.c. Josephson effect in $d$-wave superconductors,
two effects, $i.e.$ the internal phase of the pair potential
and the ZES formation,
should be taken into account.
In the previous  papers, 
we have presented a general 
formula (referred to as TK formula) for the d.c. Josephson current, 
which includes both of these effects $^{9}$ for 
anisotropic superconductors. 
This theory is based on a microscopic
basis, and the current is represented in terms of the coefficients of the
Andreev reflection $^{10-12}$.
According to  TK formula,
for a fixed phase difference between two superconductors,
the component of the Josephson current
becomes either positive or negative
depending on the injection angle of the quasiparticle
under the influence of the internal phase.
Moreover,  the maximum Josephson current $I_{C}(T)$ 
is shown to have an anomalous temperature $(T)$
dependence, that is, 
$I_{C}(T)$ becomes to be proportional to $T^{-1}$ at low temperatures
when the ZES is formed on  both electrodes $^{9,13}$.
These two effects peculiar to $d$-wave pair potential
produce a non-monotonous temperature dependence of $I_{C}(T)$
which is completely different from ordinary $s$-wave cases $^{14}$.
\par
The derivation of the previous TK formula
was based on a spatially constant pair potential model.
Although this model is suitable to obtain an analytical formula,
the spatial dependence of the pair potential should be introduced
to obtain more realistic results. 
There are mainly two effects one should take into account.
One is the suppression of the pair potential amplitude near the insulator.
Similar effect has  already been discussed
based on the quasi-classical Green's function method
at the surface of $d$-wave
superconductors $^{15-17}$.
The vanishing of
the magnitude of the $d$-wave pair potential
at the surface
has been obtained
when the angle between the normal to
the interface and the crystal axis becomes $\pi/4  + n\pi/2$,
$n$ being integers. 
The other is a possibility of spontaneous time-reversal symmetry breaking
near the insulator $^{18}$.
Under the existence of a weak $s$-wave pairing interaction, a 
subdominant $s$-wave component of the pair potential is induced
near the interface, while the bulk symmetry
remains pure $d$-wave $^{15,19}$.
Since the phase difference of the
$d$-wave and $s$-wave components is not a multiple of $\pi$,
the mixed state breaks  time-reversal symmetry. 
Under the influence of the subdominant $s$-wave component,
the ZES splits into two levels.
The possibility of  spontaneous time-reversal symmetry breaking
has been pointed out by tunneling
experiments on the surface of YBCO $^{20,21}$.
Since the anomalous temperature dependence of the d.c.
Josephson current predicted from TK formula is responsible
for the ZES formation,
it is important to check the influence of the above 
two effects on the previous results. 
Although several properties which can not be expected 
from the formula 
based on spatially constant pair potentials was obtained in 
recent theories $^{22,23,24}$, 
the influence of the
subdominant $s$-wave component on the rapid enhancement
of the Josephson current has not been clarified yet.
In this paper, we will resolve this point in detail
by extending the previous TK formula
to spatially varying potential cases. \par
In the following, we will calculate the
Josephson current in the $d$-wave superconductor / 
insulator / $d$-wave superconductor ($d/I/d$) junction. 
The assumed  spatial dependence of the pair potential 
is described as 
\begin{equation}
\label{5.5.0}
\Delta(x,\theta)
=
\left\{
\begin{array}{ll}
\Delta_{L}(x,\theta)\exp(i\varphi_{L}), & (x<0) \\
\Delta_{R}(x,\theta)\exp(i\varphi_{R}), & (x>0)
\end{array}
\right. \
\left\{
\begin{array}{ll}
\Delta_{L}(-\infty,\theta)
= \bar{\Delta}_{L}(\theta) =\Delta_{0}\cos[2(\theta-\alpha)],\\
\Delta_{R}(\infty,\theta)
= \bar{\Delta}_{R}(\theta) =\Delta_{0}\cos[2(\theta-\beta)].
\end{array}
\right.
\end{equation}
\noindent
In the above, $\theta$ and $\alpha$ ($\beta$)
denotes the injection angle of the quasiparticle, and
the angle between the normal to the interface
and the crystal axis of the left (right) superconductors, respectively
[see  Fig. 1(b) of Ref. 9(b)].
The insulator which is 
located between the two superconductors
is modeled by a $\delta$ function. 
The magnitude of the $\delta$ function denoted as $H$
determines the transparency of the junction $\sigma_{N}$.
In the case of spatially constant pair potentials,
with $\Delta_{L}(x,\theta)=\bar{\Delta}_{L}(\theta)$,
and $\Delta_{R}(x,\theta)=\bar{\Delta}_{R}(\theta)$,
the Josephson current
obtained by TK formula is  expressed  as$^{25}$
\begin{equation}
\label{5.5.1}
R_{N}I(\varphi)
= \frac{\pi \bar{R}_{N} k_{B}T}{e }
\{ \sum_{\omega_{n}}
\int^{\pi/2}_{-\pi/2}
{\rm Im} [ \mbox{\boldmath $G_{12}$}(0_{-},0_{-},\theta,i\omega_{n})
\mbox{\boldmath $G_{21}$}(0_{+},0_{+},\theta,i\omega_{n})]
t(\theta) \cos\theta d\theta \}.
\end{equation}
In the above, $t(\theta)$,
[$t(\theta) = \sigma_{N}(\theta)/(\rho_{N}^{2}(\theta)\pi^{2})]$
denotes the matrix element of the tunneling Hamiltonian
with
$\rho_{N}(\theta) = (m/\pi k_{F}\cos\theta \hbar^{2})$, where $k_{F}$ is the
magnitude of the Fermi momentum in the $d$-wave superconductor.
The quantity
$\mbox{\boldmath $G_{12}$}(0_{-},0_{-},\theta,i\omega_{n})$
($\mbox{\boldmath $G_{21}$}(0_{+},0_{+},\theta,i\omega_{n})$ )
denotes the anomalous Green's function
(conjugate of the anomalous Green's function) at
the interface with the Matsubara frequency
 $\omega_{n}=2\pi k_{B}T(n+0.5)$.
 They are  given as
\begin{equation}
\label{5.5.2}
\mbox{\boldmath $G_{12}$}(0_{-},0_{-},\theta,i\omega_{n})
=-(\frac{m}{k_{Fx}\hbar^{2}})
\frac{2\bar{\eta}_{L,+}\exp(i\varphi_{L}) }
{1 + \bar{\eta}_{L,+}\bar{\eta}_{L,-}}, \ \
\mbox{\boldmath $G_{21}$}(0_{+},0_{+},\theta,i\omega_{n})
=-(\frac{m}{k_{Fx}\hbar^{2}})
\frac{2\bar{\eta}_{R,+}\exp(-i\varphi_{R})}
{1 + \bar{\eta}_{R,+}\bar{\eta}_{R,-}}.
\end{equation}
\begin{equation}
\label{5.5.3}
\bar{\eta}_{L,\pm}=
\frac{{\rm sgn}(\omega_{n})
\mid \bar{\Delta}_{L}(\theta_{\pm}) \mid \exp(\pm i\alpha_{\pm})}
{\omega_{n} + \Omega_{n,L,\pm}}, \ \
\bar{\eta}_{R,\pm}=
\frac{{\rm sgn}(\omega_{n})
\mid \bar{\Delta}_{R}(\theta_{\pm}) \mid \exp(\mp i\beta_{\pm})}
{\omega_{n} + \Omega_{n,R,\pm}},
\end{equation}
\begin{equation}
\label{5.5.4}
\Omega_{n,R(L),\pm}={\rm sgn}
(\omega_{n}) \sqrt{\bar{\Delta}_{R(L)}^{2}(\theta_{\pm})
+ \omega_{n}^{2}}, \ \
\theta_{+}=\theta, \ \ \theta_{-}=\pi - \theta
\end{equation}
\begin{equation}
\label{5.5.5}
\exp(i\alpha_{\pm}) = \frac{\bar{\Delta}_{L}(\theta_{\pm})}
{\mid \bar{\Delta}_{L}(\theta_{\pm}) \mid }, \
\exp(-i\beta_{\pm}) = \frac{\mid \bar{\Delta}_{R}(\theta_{\pm}) \mid}
{\bar{\Delta}_{R}(\theta_{\pm})}.
\end{equation}
The quantity $R_{N}$ denotes the normal resistance
and $\bar{R}_{N}$ is expressed as
\begin{equation}
\label{5.5.6}
\bar{R}_{N}^{-1} = \int^{\pi/2}_{-\pi/2} \sigma_{N} \cos\theta d\theta, \ \
\sigma_{N}=
\frac{\cos^{2}\theta }{\cos^{2}\theta + Z^{2} }, \
Z=\frac{mH}{\hbar^{2}}
\end{equation}
\par
Let us move to the case where the pair potential has a 
spatial dependence.
Even if we take into account 
the spatial dependence of the pair potential, 
Eq. (1) does not change. Only the
anomalous Green's functions
$\mbox{\boldmath $G_{12}$}(0_{-},0_{-},\theta,i\omega_{n})$ and
($\mbox{\boldmath $G_{21}$}(0_{+},0_{+},\theta,i\omega_{n})$ )
are influenced  by the spatial dependence of the
pair potentials.
The resulting Josephson current can be
expressed as
\begin{equation}
\label{5.5.7}
R_{N}I(\varphi)= \frac{\pi \bar{R}_{N} k_{B}T}{e }
\{ \sum_{\omega_{n}}
\int^{\pi/2}_{-\pi/2} \bar{F}(\theta,i\omega_{n},\varphi)
\sigma_{N}
\cos\theta d\theta \},
\end{equation}
\begin{equation}
\label{5.5.8}
\bar{F}(\theta,i\omega_{n},\varphi)
=
\frac{4\eta_{L,+}(0_{-},\theta)\eta_{R,+}(0_{+},\theta) \exp(i\varphi)}
{ [1 + \eta_{L,+}(0_{-},\theta)\eta_{L,-}(0_{-},\theta) ]
[1 + \eta_{R,+}(0_{+},\theta)\eta_{R,-}(0_{+},\theta) ]}.
\end{equation}
with $\varphi= \varphi_{L} -\varphi_{R}$.
In the above, $\eta_{L,\pm}(0_{-},\theta)$
and $\eta_{R,\pm}(0_{+},\theta)$
are obtained by solving the following equations $^{15,16}$
\begin{equation}
\label{5.5.9}
-\frac{d}{dx} \eta_{L,+}(x,\theta)
=\frac{1}{\hbar v_{F}\cos\theta}
[\Delta_{L}^{*}(x,\theta_{+})\eta_{L,+}^{2}(x,\theta)
-\Delta_{L}(x,\theta_{+}) +2\omega_{n}\eta_{L,+}(x,\theta)]
\end{equation}
\begin{equation}
\label{5.5.10}
-\frac{d}{dx} \eta_{L,-}(x,\theta)
=\frac{1}{\hbar v_{F}\cos\theta}
[\Delta_{L}(x,\theta_{-})\eta_{L,-}^{2}(x,\theta)
-\Delta_{L}^{*}(x,\theta_{-}) +2\omega_{n}\eta_{L,-}(x,\theta)]
\end{equation}
\begin{equation}
\label{5.5.11}
\frac{d}{dx} \eta_{R,+}(x,\theta)
=\frac{1}{\hbar v_{F}\cos\theta}
[\Delta_{R}(x,\theta_{+})\eta_{R,+}^{2}(x,\theta)
-\Delta_{R}^{*}(x,\theta_{+}) +2\omega_{n}\eta_{R,+}(x,\theta)]
\end{equation}
\begin{equation}
\label{5.5.12}
\frac{d}{dx} \eta_{n,R,-}(x,\theta)
=\frac{1}{\hbar v_{F}\cos\theta}
[\Delta_{R}^{*}(x,\theta_{-})\eta_{R,-}^{2}(x,\theta)
-\Delta_{R}(x,\theta_{-}) + 2\omega_{n}\eta_{R,-}(x,\theta)].
\end{equation}
with the boundary conditions
$\eta_{L,\pm}(-\infty,\theta) = \bar{\eta}_{L,\pm}$ and
$\eta_{R,\pm}(\infty,\theta)  = \bar{\eta}_{R,\pm}$.
In the above derivations, we have assumed that
the spatial dependence of the pair potentials
$\Delta_{R}(x,\theta_{\pm})$ and
$\Delta_{L}(x,\theta_{\pm})$
can be determined independently as a semi-infinite superconductor
based on the quasi-classical Green's function method$^{15,16,23,24}$.
This assumption is valid for low transparent junctions
($\sigma_{N} \ll 1$).
\par
In the following, $I(\varphi)$ is calculated for symmetric
junction configuration, $i.e.$ cases where $\alpha=-\beta$ are satisfied.
The quantities $\Delta_{R}(x,\theta_{\pm})$ and $\Delta_{L}(x,\theta_{\pm})$
are decomposed as
\begin{equation}
\Delta_{R}(x,\theta_{\pm})
=\Delta_{R,d}(x)\cos[2(\theta \pm \alpha)] + \Delta_{R,s}(x), \
\Delta_{L}(x,\theta_{\pm})
=\Delta_{L,d}(x)\cos[2(\theta \mp \alpha)] + \Delta_{L,s}(x).
\end{equation}
Since the temperature dependence of the pair potential amplitude
is assumed to follow ordinary BCS relation,
the $T_{C}$'s for $s$-wave and $d$-wave components (referred
to as $T_{s}$ and $T_{d}$) directly correspond to
the magnitude of the attractive interaction  for the two component.
We will consider two cases;
(1) $\Delta_{L,d}(-x)=\Delta_{R,d}^{*}(x)$,
$\Delta_{L,s}(-x)=\Delta_{R,s}^{*}(x)$
and
(2) $\Delta_{L,d}(-x)=\Delta_{R,d}(x)$, 
$\Delta_{L,s}(-x)=\Delta_{R,s}(x)$ with positive $x$. 
In general, 
the relative phase between the $s$ and $d$-wave components 
of the pair potentials, $i.e.$, $\Delta_{R(L),d}(x)$ 
and $\Delta_{R(L),s}(x)$, is not a multiple of $\pi$, 
which implies that the time reversal symmetry is broken.
The extreme case is $\alpha=\pm \pi/4$, where the relative phase
becomes $\pm \pi/2$.
In Figs. 1 and 2,
the temperature dependence of $I_{C}(T)$ is plotted for the case (1).
As a reference, the results for the spatially constant pair potential case
are plotted as curves A in each figures.
The transition temperature for the $s$-wave pair potential $T_{s}$
without $d_{x^{2}-y^{2}}$-wave pair potential 
is chosen as $T_{s}=0$ for curves B in Figs.1 and 2,
$T_{s}=0.09T_{d}$ for curve C in Fig. 2, and 
$T_{s}=0.17T_{d}$ for curve C in Fig.1 and 
curve D in Fig. 2. 
For these parameters,
$s$-wave component is subdominantly induced only  near the interface.
For $\alpha=0.1\pi$,
the non-monotonous temperature dependence of $I_{C}(T)$
is obtained even if the spatial dependence of the $d$-wave component
of the pair potential
is taken into account (see  curve B in Fig. 1).
The deviation  of  $I_{C}(T)$ from curve A
is much more drastic for
$\alpha=\pi/4$ (see curve B in Fig. 2)
where the suppression of the pair potential is most significant $^{13,22}$.
The result shows that 
although the amplitude of the current is reduced,
the qualitative feature in the  $I_{C}(T)$
does not change seriously as compared to curve A $^{13,22}$.
On the other hand, when the $s$-wave component, 
which breaks the time reversal symmetry, is induced,
the enhancement of $I_{C}(T)$ at low
temperatures is significantly suppressed (see curve C in Figs.
1 and 2). 
With the increase of the magnitude of the induced $s$-wave component,
the suppression of $I_{C}(T)$ becomes much more significant
(see curve D in Fig. 2).
These features are intuitively explained as follows. 
Under the existence of the $s$-wave component which breaks the time
reversal symmetry,  ZEP 
in local density of states (LDOS) is split into two, and
LDOS at zero energy is reduced.
At the same time, 
the magnitude of the anomalous Green's functions
$\mbox{\boldmath $G_{12}$}(0_{-},0_{-},\theta,i\omega_{n})$ and
($\mbox{\boldmath $G_{21}$}(0_{+},0_{+},\theta,i\omega_{n})$ )
for low energies decreases as compared to that for $T_{s}=0$. 
Thus the enhancement of $I_{C}(T)$
is suppressed.
\par
In Fig. 3, the effect of the relative phase of the induced $s$-wave
component between the left and the right superconductors is discussed
for symmetric junction configuration ($\alpha=-\beta=\pi/4$).
In the case of (1),
$\Delta_{R}(x,\theta)$ and
$\Delta_{L}(-x,\theta)$ are given as
$[\Delta_{d}(x)\sin(2\theta) - i\Delta_{s}(x)]$ and
$[-\Delta_{d}(x)\sin(2\theta) + i\Delta_{s}(x)]$, respectively 
for $x>0$. 
The relative  phase between
$\Delta_{R}(x,\theta)$ and
$\Delta_{L}(x,\theta)$ is always $\pm \pi$ independent of the
magnitude of $\Delta_{s}(x)$.
In this case, the position of the 
free energy minima of the junction is always 
located at $\varphi= \pm \pi$ independent of the temperature
and $I_{C}(T)$ increases monotonically with the decrease of the
temperature (see curve B in Fig. 3).
On the other hand, for case (2),
$\Delta_{R}(x,\theta)$ and
$\Delta_{L}(-x,\theta)$ are given as
$[\Delta_{d}(x)\sin(2\theta) + i\Delta_{s}(x)]$ and
$[-\Delta_{d}(x)\sin(2\theta) + i\Delta_{s}(x)]$, respectively 
for $x>0$. 
The relative  phase between
$\Delta_{R}(x,\theta)$ and
$\Delta_{L}(x,\theta)$ deviates from  $\pm \pi$ depending on the
magnitude of $\Delta_{s}(x)$.
The Josephson current carried by the
induced $s$-wave component of the pair potential
flows in the  direction opposite to that of the
$d$-wave component. Consequently, the position of the
free energy minima of this junction changes
from $\pi$ to 0 with the decrease of the temperature. This
fact arouses the
 non-monotonous temperature dependence of $I_{C}(T)$
(see curve C in Fig. 3). \par
In this paper, we have discussed 
the effect of the spatial 
dependence of the pair potential on the d.c. Josephson current 
in  $d/I/d$ junctions. 
Although the suppression of the pair potential
near the insulator affects  the magnitude$^{22}$ of $I_{C}(T)$, 
the temperature dependence of $I_{C}(T)$ 
is similar to that in spatially constant pair potential cases. 
On the other hand, the inducement of the 
subdominant $s$-wave component
which breaks the time reversal symmetry
has a large influence on the qualitative features of the
Josephson effect.
Especially, the enhancement of the $I_{C}(T)$ at low temperatures
due to the ZES is strongly suppressed.
Also it is a remarkable fact that
the relative phase between the induced $s$-wave components of the
pair potentials
of the left and right superconductors
can be distinguished through the temperature dependence of $I_{C}(T)$.
\par
One of the serious problems in the field of the  Josephson effect in
high-$T_{c}$ superconductors is the lack of the consistency between
theories and experiments.
By comparing the present results with experimental data in detail, 
we expect to obtain consistency with the microscopic 
information of the paring interaction strength. 
Also it is an interesting problem to clarify how the
roughness at the interface $^{13,22}$ influences 
the subdominant pair potential and  the d.c. Josephson effect. \par
\vspace{0.5cm}
We would like to thank to M. Koyanagi, K. Kajimura,
K. Nagai, M. Matsumoto, Y. Ohhashi, Y. Nagato,
M. R. Beasley, J. Mannhart, and H. Hilgenkamp
for stimulating discussions.
This work is supported by a Grant-in-Aid for Scientific
Research in Priority Areas
''Anomalous metallic state near the Mott transition''
and ''Nissan Science Foundation''.
The computational aspect of this work has been done for the
facilities of the Supercomputer Center, Institute for Solid State Physics,
University of Tokyo and the Computer Center, Institute for Molecular
Science, Okazaki National Research Institute.
\newpage

\noindent
$^{+}$Permanent Adress: Electrotechnical Laboratory, Tsukuba,
Ibaraki 305-9568, Japan
\par
\noindent
$^{1}$D.J. Scalapino, Physics Report. {\bf 250}, 329 (1995).
\par
\noindent
$^{2}$D. J. Van Harlingen, Rev.~Mod.~Phys. {\bf 67}, 515 (1995).
\par
\noindent
$^{3}$C. R. Hu,  Phys.~Rev.~Lett. {\bf 72}, 1526 (1994);
J. Yang and C. R. Hu, Phys. Rev. B, {\bf 50}, 16766, (1995).
\par
\noindent
$^{4}$Y. Tanaka and S. Kashiwaya,
Phys. Rev. Lett., {\bf 74}, 3451 (1995);
Phys. Rev. B, {\bf 53}, 9371, (1996).
\par
\noindent
$^{5}$S. Kashiwaya $et$.$al$.
Phys. Rev. B, {\bf 51}, 1350, (1995); $ibid.$,
{\bf 53}, 2667, 1996.
\par
\noindent
$^{6}$J. Geerk $et.$ $al.$,
Z.~Phys.~B {\bf 73}, 329, (1988).
\par
\noindent
$^{7}$I. Iguchi  and Z. Wen, 1991,
Physica~C {\bf 178}, 1 (1991).
\par
\noindent
$^{8}$ L. Alff $et.al.$,
Phys. Rev. B, {\bf 55}, 14757, (1997);
S. Ueno $et.al.$,
to be published in J. of Phys. Chem. Solids (1998).
\par
\noindent
$^{9}$Y. Tanaka and S. Kashiwaya,
(a)Phys. Rev. B, {\bf 53}, 11957, (1996); (b)$ibid.$, {\bf 56}, 892, (1996).
\par
\noindent
$^{10}$A. F. Andreev, Zh.~Eksp.~Teor.~Fiz. {\bf 46}, 1823 (1964).
[Trans. Sov.~Phys.~JETP, {\bf 19}, 1228]. \par
\noindent
$^{11}$A. Furusaki and M. Tsukada,
Solid~State~Commun. {\bf 78}, 299, (1991).
\par
\noindent
$^{12}$Y. Tanaka, Phys. Rev. Lett., {\bf 72}, 3871 (1994).
\par
\noindent
$^{13}$ Yu. S. Barash $et.$ $al.$,
Phys.~Rev.~Lett. {\bf 77}, 4070 (1996).
\par
\noindent
$^{14}$V. Ambegaokar and A. Baratoff,  Phys.~Rev.~Lett.
{\bf 10}, 486 (1963).
\par
\noindent
$^{15}$M. Matsumoto, H. Shiba, J. Phys. Soc. Jpn., {\bf 64},
3384, (1995); {\bf 64}, 4867 (1995).
\par
\noindent
$^{16}$Y. Nagato $et.$ $al.$,
J.~Low.~Temp.~Phys., {\bf 93}, 33 (1993);
Y. Nagato, K. Nagai, Phys. Rev. B {\bf 51},
16254 (1995).
\par
\noindent
$^{17}$L.J. Buchholtz $et.$ $al.$,
J.~Low~Temp.~Phys. {\bf 101}, 1097 (1995).
\par
\noindent
$^{18}$
M. Sigrist $et.$ $al.$, Phys. Rev. B {\bf 53}, 2835 (1996).
\par
\noindent
$^{19}$M. Fogelstr\"{o}m $et.$ $al.$,
Phys.~Rev.~Lett. {\bf 79}, 281 (1997).
\par
\noindent
$^{20}$M. Covington $et.$ $al.$,
Phys.~Rev.~Lett. {\bf 79}, 277 (1997).
\par
\noindent
$^{21}$ S. Kashiwaya $et.al.$,
to be published in J. of Phys. Chem. Solids (1998). \par
\noindent
$^{22}$ H. Burkhardt, {\it Quasiclassical Methods in 
in Superconductivity and Superfluidity}, editied by D.A. Rainer and 
J.A. Sauls (Springer, Heidelberg, 1998).  
\par
\noindent
$^{23}$ M. Fogelstr\"{o}m $et.$ $al.$,
to be published in Physica C (1998), cond-mat(9709120).
In this paper, $d_{xy}$($B_{2g}$) symmetry
is chosen as the subdominant pair potential. \par
\noindent
$^{24}$ S. Yip, J. Low. Temp. Phys. 
{\bf 109}, 547 (1997). \par
\noindent
$^{25}$A.J. Millis, $et$. $al.$, Phys. Rev. B {\bf 38},
4504 (1988). \par
\newpage
\begin{figure}
\caption{Maximum Josephson current $I_{C}(T)$
in $d/I/d$ junction
with $\alpha=-\beta=0.1\pi$ plotted as a function of
temperature with $Z=10$.
A: Step function model,
B: Self-consistently determined pair potential without
$s$-wave component, $T_{s}=0$
C: Self-consistently determined pair potential with
$s$-wave component. $T_{s}=0.17T_{d}$.}
\label{f1}
\end{figure}

\begin{figure}
\caption{Maximum Josephson current $I_{C}(T)$
in $d/I/d$ junction
with $\alpha=-\beta=0.25\pi$ plotted as a function of
temperature with $Z=10$.
A: Step function model,
B: Self-consistently determined pair potential without
$s$-wave component, $T_{s}=0$,
C: Self-consistently determined pair potential with
$s$-wave component. $T_{s}=0.09T_{d}$.
D: Self-consistently determined pair potential with
$s$-wave component. $T_{s}=0.17T_{d}$.}
\label{f2}
\end{figure}

\begin{figure}
\caption{Maximum Josephson current $I_{C}(T)$
in $d/I/d$ junction
with $\alpha=-\beta=0.25\pi$ plotted as a function of
temperature with $Z=10$.
A: Step function model,
B: Self-consistently determined pair potential
under the coexistence of the
$s$-wave component, with
$\Delta_{L,d}(-x) = \Delta_{R,d}^{*}(x) $,
$\Delta_{L,s}(-x) = \Delta_{R,s}^{*}(x) $.
C: Self-consistently determined pair potential with
$s$-wave component with
$\Delta_{L,d}(-x)=\Delta_{R,d}(x)$,
$\Delta_{L,s}(-x)=\Delta_{R,s}(x)$.
In B and C, $T_{s}$ is chosen as
$T_{s}=0.09T_{d}$.}
\label{f3}
\end{figure}

\end{document}